\documentclass[3p,twocolumn]{elsarticle}
\usepackage{color}
\usepackage{amssymb}
\journal{Astroparticle Physics}
\begin{document}
\begin{frontmatter}


\title{A detailed study on the range fluctuations of $10^{11}$eV to $10^{18}$eV muons in water and the fluctuations of the Cherenkov light yields due to the accompanied cascade showers initiated by muons\\ Part1: High Energy Muons}
\author[1]{Y.Okumura}
\author[1]{N.Takahashi}
\author[2,3]{A.Misaki}
\address[1]{Graduate School of Science and Technology Hirosaki University, Hirosaki, 036-8561, Japan}
\address[2]{Innovative Research Organization, Saitama University, Saitama, 338-8570, Japan}
\address[3]{The Institute for China-Japan Culture Study, Mitaka, Tokyo, 113-0004, Japan}

\begin{abstract}
 The validity of our Monte Carlo simulation procedure (\textit{the time sequential procedure}) had been verified by the corresponding analytical procedure which is independent of our method methodologically. Also, the results obtained by our procedure are compared with those  obtained by the different Monte Carlo simulation procedure (\textit{the} $v_{cut}$ \textit{procedure}) which has been exclusively utilized by  different authors and the agreement between ours and theirs are found to be good. By utilizing our Monte Carlo procedure, the validity of which is guaranteed in two different procedures, we investigate not only  the fluctuation of high energy muons themselves but also the fluctuation of the various quantities related to the energy losses  by the muons, which are difficult to obtain by \textit{the} $v_{cut}$ \textit{procedure}. Namely, we obtain the fluctuation on energy losses of  muons in the present paper(Part1). In a subsequent paper (Part2), based on the idea developed in Part1, we will examine the correlation between the primary energies of  muons and their energy deposits and the correlation between the energy deposits of the muons concerned and their Cherenkov light yields in the KM3 detector in detail.
\end{abstract}
\begin{keyword}
High energy muon \sep  
Muon range fluctuation \sep 
Muon energy loss \sep
Muon propagation simulation
\end{keyword}
\end{frontmatter}
%
%
\section{Introduction}
\label{intro}
In the KM3 detectors deployed in the Antarctic, the ocean and  lakes \cite{IceCube, Antarctic, Mediterranean, Baikal}, it is most important  to decide the energies of high energy muons due to  neutrino interactions  from the experimental and technical point of view.
In the energy determination of high energy muons, there are two indispensable problems to be carefully examined: the first is to examine the fluctuation of the traversed lengths of  muons and the second is the fluctuation on the Cherenkov light yields initiated by both the muons concerned and their daughters' electromagnetic cascade showers.
 In the present paper (Part1), we studied the range fluctuation of high energy muons in detail and in a subsequent paper (Part2), we will study the fluctuation on the Cherenkov light yields of the electromagnetic cascade showers which influence decisively over the energy determination of the high energy muons in detail.

The studies on the range fluctuations of high energy muons have a long history \cite{Mando}-\cite{Allkofer} even before the appearance of the high performance of electronic computers, but the recent big progress of the computers makes it possible to treat fluctuation problems on high energy muons in a more quantitative manner using the Monte Carlo method.
 As far as the treatment of the range fluctuation of high energy muons by the 
Monte Carlo method is concerned, there exist two independent 
approaches. 
The first one is  the procedure of splitting the radiative energy loss into two terms, a continuous "soft" term for $v < v_{cut}$ and a stochastic "hard" term for $v_{cut} < v < 1$ in the equation for energy loss (see, Eq.(\ref{dEdx_rad})) \cite{Lipari}-\cite{Bottai}. Hereafter, we will simply refer to it as \textit{the} $v_{cut}$ \textit{procedure}. The introduction of $v_{cut}$ is for the purpose of applying the Monte Carlo methods to the  "hard" term and its value is carefully chosen to save computational time.
 The other is {\it the time sequential procedure} developed by us in which the interaction points of the muons and their dissipated energies are directly determined \cite{Takahashi, Takahashi2} and here, all the processes are treated in the stochastic manner without the introduction of $v_{cut}$.
These two methods are independent of each other, but are logically equivalent, giving the same results as for the muons' behaviors (see Figures \ref{fig:Lipari} to 6). 
Here, it should be noticed that in KM3 detectors, the energy determination of high energy muons are carried out by measuring Cherenkov light yields which are produced by not only the muons themselves but also by the electromagnetic cascade showers induced by the muons concerned. Then it should be emphasized that the largest Cherenkov light yields are occupied by those generated by  electromagnetic cascade showers and the Cherenkov light yields produced by the original muons can be neglected for muon energies above $\sim 10^{14}$ eV (See detail, in the subsequent paper[27]).
These cascade showers are generated from the stochastic processes, such as bremsstrahlung, direct electron pair production and photo nuclear interaction which are initiated by muons concerned. 
In the present paper, we discuss the behaviors of high energy muons in water(ice) and the problems around Cherenkov light yields are treated in a subsequent paper. 
\section{Range fluctuation of the (ultra-) high energy muons and individual behavior of the muons}
\label{sec:2}
\subsection{The physical meaning of "no fluctuation"}
\label{sec:2.1}
\ The average energy loss by high energy muons is usually 
described as,
\begin{equation}
\label{dEdx}	               
- \frac{dE}{dx}=a\left(E\right)+b\left(E\right)\cdot E,
\end{equation}
where {\it a} is the term due to ionization which is free from fluctuation and {\it b} is the term due to stochastic processes which may be origins of fluctuations. The latter is divided into three parts. Namely,
\begin{equation}
\label{bEmu}	               
b\left(E\right)=b_{b}\left(E\right)+b_{d}\left(E\right)+b_{n}\left(E\right),
\end{equation}
where ${\it b}_{b}$, ${\it b}_{d}$ and ${\it b}_{n}$ are the corresponding terms due to bremsstrahlung, direct electron pair production and photo nuclear interaction, respectively.
 In the treatment of the average energy loss, each {\it b}-term is defined as,
\begin{equation}
\label{intb}	               
b_{i}\left(E\right)=\frac{N}{A}\int_{v_{min}}^{v_{max}}v\left[\frac{d\sigma_{i}\left(v,E\right)}{dv}\right]\cdot dv,
\end{equation}
where $v_{max}$ and $v_{min}$ are the maximum and the minimum fractional energies due to their kinematical limits. Here, N and A denote the Avogadro number and the Atomic number of the medium. The physical meaning of Eq.(\ref{dEdx}) is that the muons concerned dissipate energy in the form of the average uniquely, being defined by Eq.(\ref{intb}). 

\begin{figure}[!t]
\begin{center}
\resizebox{0.425\textwidth}{!}{\includegraphics{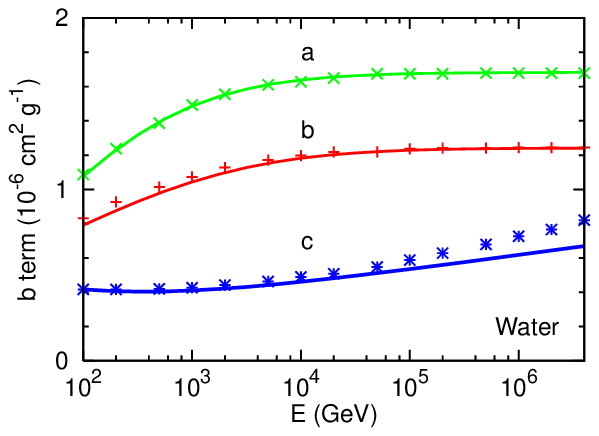}}
\caption{\textit{b}-terms due to different stochastic processes. a, b and c denote direct electron pair, bremsstrahlung and photo nuclear interaction, respectively. The lines denote ours, while symbols are due to Klimushin et.al.}
\label{fig:B-TERM_W}            
\resizebox{0.45\textwidth}{!}{\includegraphics{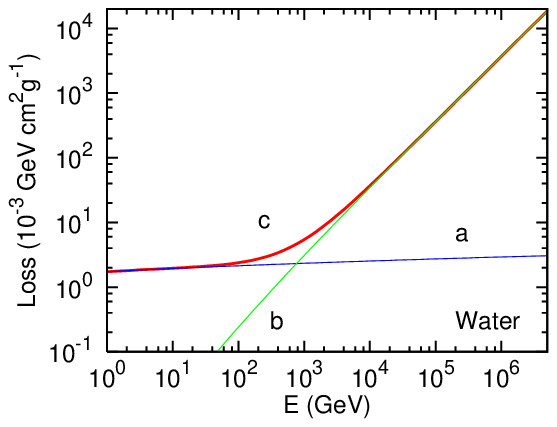}}
\caption{The relation between {\it a}-term and {\it b}-term in water.}
\label{fig:dEdx}	            
\resizebox{0.45\textwidth}{!}{\includegraphics{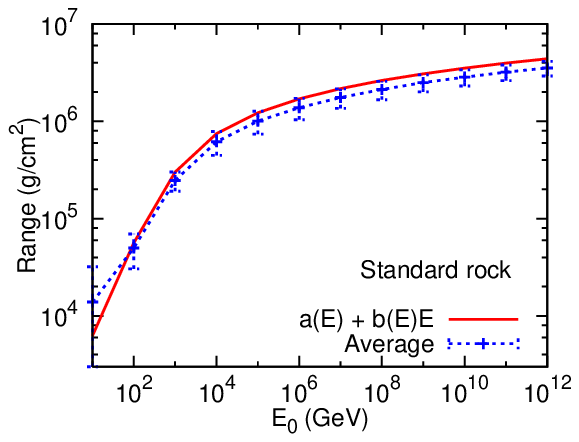}}
\caption{The range energy relation between the case without fluctuation and the case with fluctuation in standard rock. The uncertainty bar denote the standard deviation.}
\label{fig:RangeSR}	            
\end{center}
\end{figure}

Namely, fluctuations are not included in Eq.(\ref{intb}).
In Figure \ref{fig:B-TERM_W}, we give the {\it b}-terms due to different processes in water.
 In Figure \ref{fig:dEdx}, we give the relation between $a(E)$ and $b(E)$. As the \textit{b}-terms are essentially of a stochastic character, it is seen from the figure that the stochastic processes become effective above $\sim 1$TeV($10^{12}$eV). Therefore, we must treat the muon's behavior in a stochastic manner above $\sim$ 1 TeV. Below $\sim 1$TeV we may treat muon's behavior in the non-stochastic manner.
Then, the range of the muon is uniquely determined by Eq.(\ref{Rdx}).
\begin{equation}
\label{Rdx}	                    
R=\int_{E_{min}}^{E_{0}}\frac{dE}{-\frac{dE}{dx}},
\end{equation}
where $E_{min}$ is the minimum energy for observation and $E_{0}$ is the primary energy of the muon.
\ Throughout the present paper, $E_{min}$ denotes the minimum energy among the energies for observation ($E_{obs}$) and is taken as $1$GeV ($10^{9}$eV). Thus, $R$ defined by Eq.(\ref{Rdx}) gives the effective range of the muon without fluctuation. 

 In Figure \ref{fig:RangeSR}, we give the effective range defined by Eq.(\ref{Rdx}) together with our average ranges of the muons in which the fluctuation effects are exactly taken into account.(see discussion in the later sections). It should be noticed from the figure that the effective ranges without fluctuation are different from the average range of the muons in which the fluctuation is considered, on which Lipari and Stanev\cite{Lipari} already pointed out\footnote{What Lipari and Stanev pointed out had been confirmed by J.Nishimura in the analysis of  intensity-depth relation of high energy muons. See page 111 Table 22. \cite{Nishimura2}.}. 
 Really, the real average ranges are smaller than those of effective range beyond one standard deviation above $10^{13}$ eV as shown in Figure \ref{fig:RangeSR}.
%
\subsection{Physical quantities with fluctuation}
\label{sec:2.2}
\ In {\it the $v_{cut}$ procedure}, many authors \cite{Lipari}- \cite{Bottai} divide all stochastic processes into two parts, namely, the continuous part and radiative part in their Monte Carlo simulation in order to consider the fluctuation in both the range and energies of the muons and introduce $v_{cut}$ to save time for computation, while we treat all stochastic processes as exactly as possible without the introduction of $v_{cut}$ in \textit{the time sequential procedure} \cite{Takahashi}\cite{Takahashi2} and the essential structure for calculation is described in the Appendix. 
The validity of our Monte Carlo method had been already checked by the analytical method based on the diffusion equation \cite{Misaki} which is methodologically independent of the Monte Carlo procedure. Furthermore, we check the validity of our method, comparing our results with the corresponding results by \textit{the $v_{cut}$ procedure}.
%
\subsubsection{The comparison of our results with others}
\label{sec:2.2.1}
The survival probability for high energy muon in our method is defined as,
\begin{equation}
\label{PEXE}	                           
P\left(E_{obs},X,E_{p}\right)=\frac{N_{through}\left(E_{obs},X\right)}{N_{sample}\left(E_{p}\right)},
\end{equation}
where $E_{p}$, $X$, and $E_{obs}$, denote the energy of the primary muon, the point of observation and the minimum energy of the muon at the point of observation, respectively. $N_{sample}\left(E_{p}\right)$ denotes the total sampling number of muons and $N_{through}\left(E_{obs},X\right)$ denotes the number of muons concerned with energies above $E_{obs}$ which pass through the observation point X.
\begin{figure}[!t]
\begin{center}
\resizebox{0.45\textwidth}{!}{\includegraphics{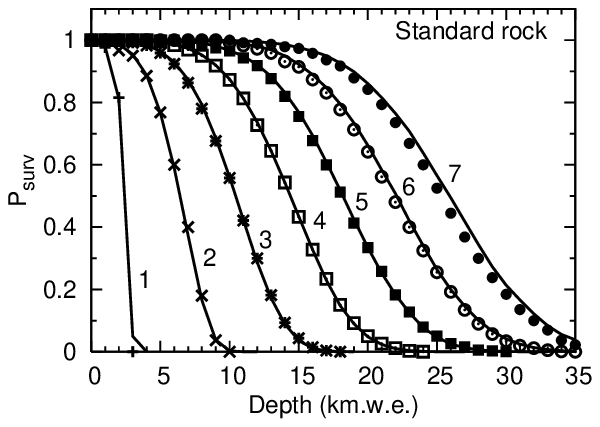}}		
\caption{The comparison of our result with that of Lipari and Stanev\cite{Lipari}. The survival probabilities of muons of energy from $1$ TeV to $10^{6}$ TeV.  The numerical figures attached each curve denote the primary energies. Curves labels correspond to following set of primay energies of muon: (1)$1$TeV, (2)$10$TeV, (3)$10^{2}$TeV, (4)$10^{3}$TeV, (5)$10^{4}$TeV, (6)$10^{5}$TeV, (7)$10^{6}$TeV. Symbols are due to Lipari and Stanev and curves are due to ours.}
\label{fig:Lipari}
\resizebox{0.45\textwidth}{!}{\includegraphics{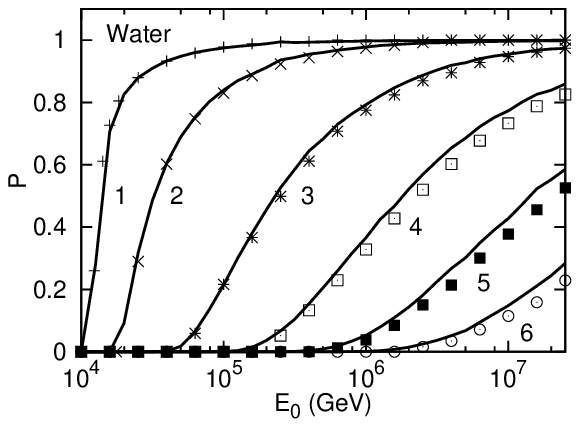}}	  
\caption{The comparison of our results with that of Klimushin et al\cite{Klimushin}.
The continuous lines are obtained by us, while symbols are readout  from those by Klimushin et al for primary energies from $10^{13}$ eV to $3\times 10^{16}$ eV. The numerical figures attached each curve denote the threshold energy is 10 TeV. Curves labels correspond to following set of depths: (1)1.15km, (2)3.45km, (3)8.05km, (4)12.65km, (5)17.25km, (6)21.39km.}
\label{fig:Buga}
\end{center}
\end{figure}
\begin{figure}[!t]
\begin{center}
\resizebox{0.45\textwidth}{!}{\includegraphics{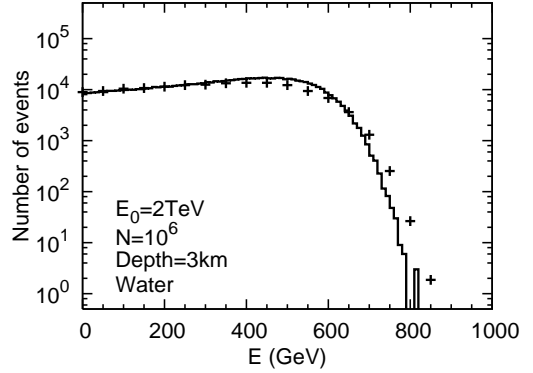}}	    
\caption{The comparison of our results with that of Kudryavtsev\cite{Kudryavtsev}. Energy spectrum of muons with initial energy of $2$ TeV at $3$ km. The line denotes ours, while symbols are due to Kudryavtsev}
\label{fig:ES3km}
\end{center}
\end{figure}

We compare our results by {\it the time sequential procedure} with the different authors' results by {\it the $v_{cut}$ procedure} in the following. Lipari and Stanev\cite{Lipari} give the survival probabilities as the functions of the depths for $1$ TeV to $10^{6}$ TeV  incident muons in water and partly standard rock, the minimum energy of which is taken as 1 GeV. We obtain the corresponding results by {\it the time sequential procedure} and compare our results with Lipari and Stanev's in Figure \ref{fig:Lipari}. 

Also, Klimushin et al\cite{Klimushin} give the survival probabilities for primary energy of $10^{13}$ eV to $3 \times 10^{16}$ eV. We obtain the corresponding results to them and compare our  results with their results in Figure \ref{fig:Buga}. 

Furthermore, Kudryavtsev\cite{Kudryavtsev} gives the energy spectrum of the muon due to primary energy of $2$ TeV at $3$ km in water. We compare our results with his in Figure \ref{fig:ES3km}.

Thus, rather nice agreements between ours and the results by different authors shown in Figures \ref{fig:Lipari} to \ref{fig:ES3km}
 are found to guarantee the validity of our method, taking into account  the slight difference in their differential cross sections such as the bremsstrahlung, direct electron pair production and photo nuclear interaction by different authors and the slight difference in their procedures, in addition to the confirmation of the validity of our method by the analytical method.
Here, we briefly mention to the essential difference between \textit{the $v_{cut}$ procedure} and \textit{the time sequential procedure}.
 In \textit{the $v_{cut}$ procedure}, one writes down the fundamental equation in the following,

\begin{eqnarray}
\label{dEdx_rad}                
\frac{dE}{dx} &=&\left[\frac{dE}{dx}\right]_{soft}+\left[\frac{dE}{dx}\right]_{hard} \nonumber \\
&=&\frac{N}{A}E\int_{0}^{v_{cut}}dv\cdot v\frac{\sigma\left(v,E\right)}{dv}
\nonumber \\
&+&\frac{N}{A}E\int_{v_{cut}}^{1}dv\cdot v\frac{d\sigma\left(v,E\right)}{dv},
\end{eqnarray}

As Eq.(\ref{dEdx_rad}) expresses the average energy loss of the muons, this method gives the same result concerning the energy loss of the muons concerned, irrespective of the $v_{cut}$. In Eq.(\ref{dEdx_rad}), the Monte Carlo procedure is carried out in the second term and the difference in $v_{cut}$ gives the different shape of the energy losses.
The only and essential difference between \textit{the} $v_{cut}$ \textit{procedure} and \textit{the time sequential procedure} lies in that these two methods may give different shapes of the emitted energies of the muons  even if their total dissipated energies are same. The difference in the shapes of the emitted energies between these two methods greatly influences their shapes of the Cherenkov light yields and their magnitude, which, in turn, may give the different estimation on the muon's energies concerned.
We will examine carefully Cherenkov light yields problems in detail in a subsequent paper(Part2)
 
\begin{figure}[!t]
\begin{center}
\resizebox{0.45\textwidth}{!}{\includegraphics{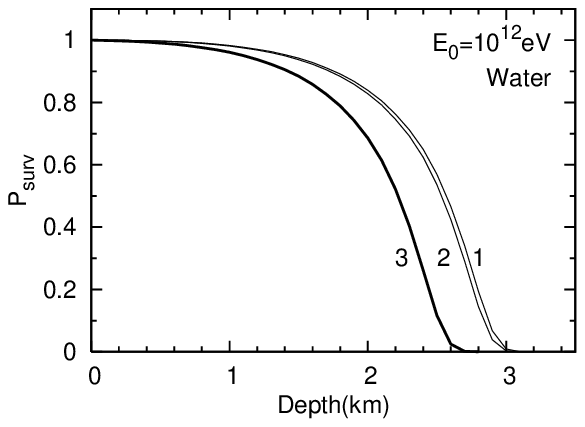}}	        
\caption{The survival probabilities for $10^{12}$eV muon.
Curves labels correspod to following set of cutoff energies:
(1)$10^{9}$eV, (2)$10^{10}$eV, (3)$10^{11}$eV.}
\label{fig:SP12}
\resizebox{0.45\textwidth}{!}{\includegraphics{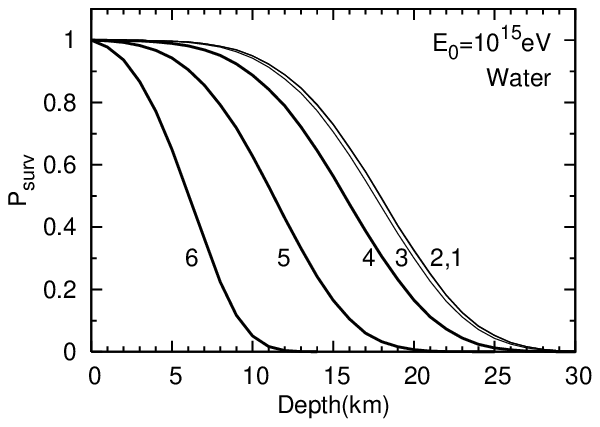}}	        
\caption{The survival probabilities for $10^{15}$eV muon.
Curves labels correspod to following set of cutoff energies:
from (1)$10^{9}$eV to (6)$10^{14}$eV.}
\label{fig:SP15}
\resizebox{0.45\textwidth}{!}{\includegraphics{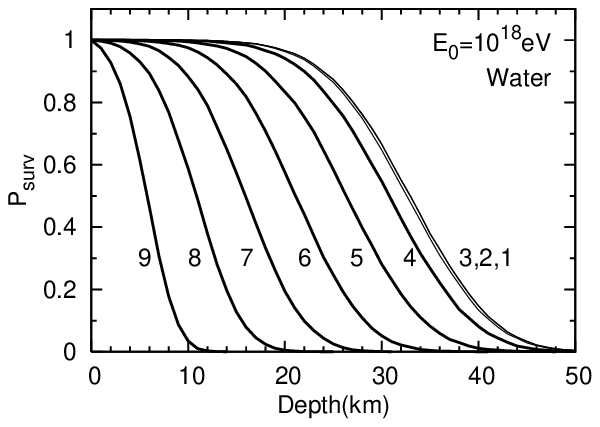}}	        
\caption{The survival probabilities for $10^{18}$eV muon.
Curves labels correspod to following set of cutoff energies:
from (1)$10^{9}$eV to (9)$10^{17}$eV.}
\label{fig:SP18}
\end{center}
\end{figure} 
%
\subsubsection{Survival probabilities and  their $v_{cut}$ energy spectra at different observable depths}
\label{sec:2.2.2}
\begin{figure}[!t]
\begin{center}
\resizebox{0.45\textwidth}{!}{\includegraphics{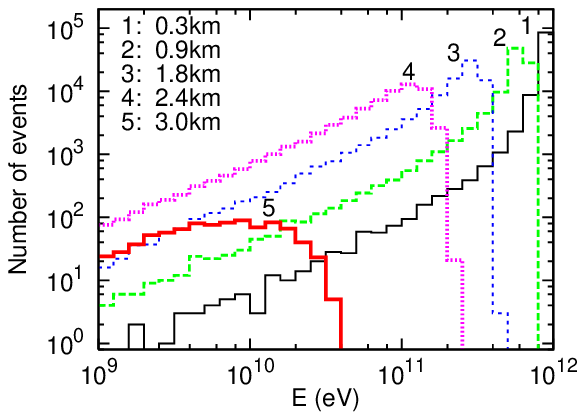}}       
\caption{Energy spectrum in water at the different depths, initiated by $10^{12}$eV muons.}
\label{fig:ES12}
\resizebox{0.45\textwidth}{!}{\includegraphics{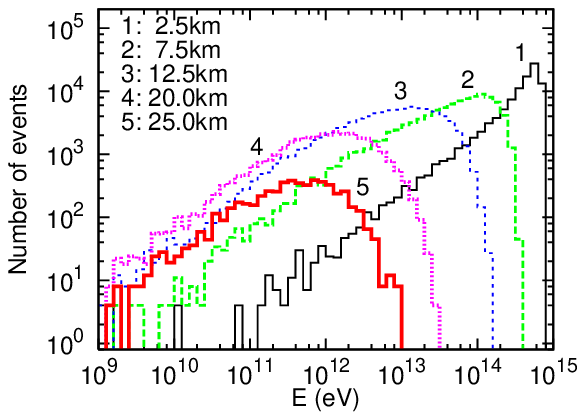}}       
\caption{Energy spectrum in water at the different depths, initiated by $10^{15}$eV muons.}
\label{fig:ES15}
\resizebox{0.45\textwidth}{!}{\includegraphics{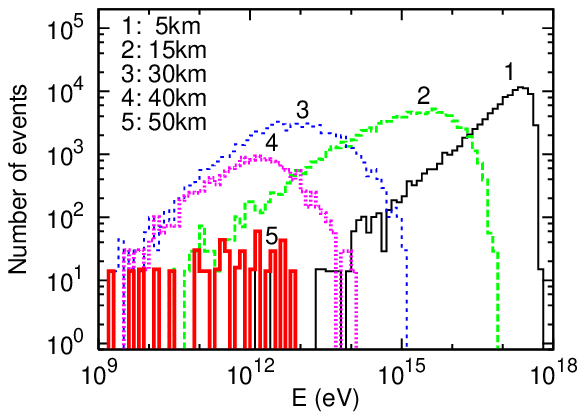}}       
\caption{Energy spectrum in water at the different depths, initiated by $10^{18}$eV muons.}
\label{fig:ES18}
\end{center}
\end{figure}
In Figures \ref{fig:SP12} to \ref{fig:SP18}, we give the survival probabilities for different observable energies with primary energies of $10^{12}$eV, $10^{15}$eV and $10^{18}$eV, respectively.  In Figure \ref{fig:SP12}, we give the observation energies $10^{9}$eV, $10^{10}$eV and $10^{11}$eV, respectively. In Figure \ref{fig:SP15}, we give them $10^{9}$eV, $10^{10}$eV, $10^{11}$eV, $10^{12}$eV, $10^{13}$eV and $10^{14}$eV, respectively. In Figure \ref{fig:SP18}, we give them, $10^{9}$eV to $10^{17}$eV, respectively. Each sampling numbers in Figure \ref{fig:SP12} to \ref{fig:SP18} are 100,000. It is seen from the figures that the survival probabilities do not decrease remarkably as their primary energies increase.

In Figures \ref{fig:ES12} to \ref{fig:ES18}, we give the differential energy spectra for primary energies, $10^{12}$eV, $10^{15}$eV and $10^{18}$eV at different depths, respectively. It is seen from the figures that the energy spectrum at the initial stage are of delta-function and they shift as the mountain-like deforming their shape in the intermediate stage and finally, 
they disappear in deeper depth for their threshold energy (1 GeV).
Each sampling number in Figures 10 to 12 is 100,000.

\subsubsection{Range Distributions of Muons}
\begin{figure}[!t]
\begin{center}
\resizebox{0.45\textwidth}{!}{\includegraphics{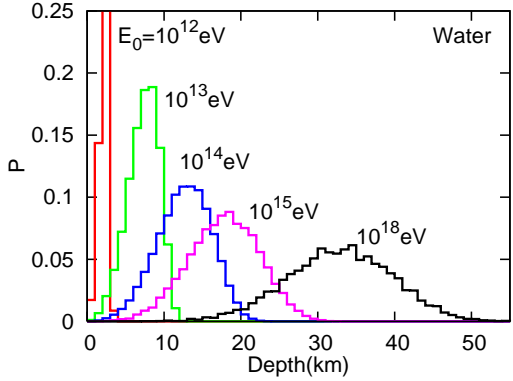}}	        
\caption{Range distributions for $10^{12}$eV to $10^{18}$eV muons in water. The minimum observation energies are taken as $10^{9}$ eV. Each sampling number is 100,000.}
\label{fig:RF131518}
\end{center}
\end{figure}
\begin{figure}[!t]
\begin{center}
\resizebox{0.45\textwidth}{!}{\includegraphics{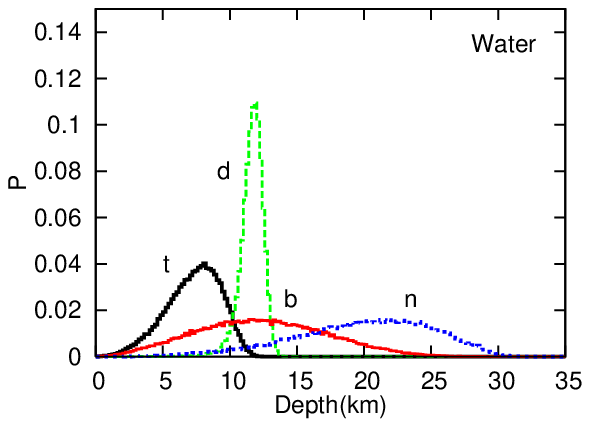}}	            
\caption{Hypothetical range distributions in water for $10^{13}$eV muons together with the real range distribution.}
\label{fig:RF13Sp}
\resizebox{0.45\textwidth}{!}{\includegraphics{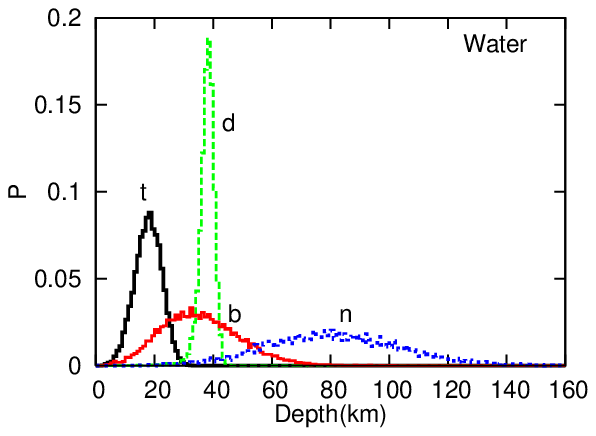}}	            
\caption{Hypothetical range distributions in water for $10^{15}$eV muons together with the real range distribution.}
\label{fig:RF15Sp}
\resizebox{0.45\textwidth}{!}{\includegraphics{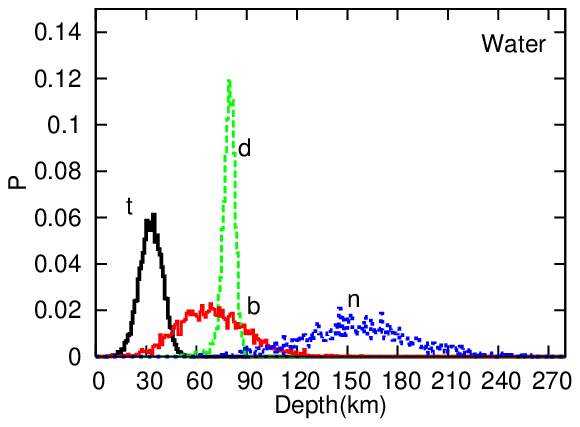}}	            
\caption{Hypothetical range distributions in water for $10^{18}$eV muons together with the real range distribution.}
\label{fig:RF18Sp}
\end{center}
\end{figure}
\ All processes, such as bremsstrahlung, direct electron pair production and photo nuclear interaction are stochastic and, therefore, one cannot neglect their fluctuation essentially. Muons penetrate through  matter as the results of the competition effect between bremsstrahlung, direct electron pair production and photo nuclear interaction. The logic on their penetration is described in the Appendix.

In Figure \ref{fig:RF131518}, we give $P\left(R;E_{0}\right)$, the probabilities for the range distribution in water with primary energies, $10^{13}$eV, $10^{15}$eV and $10^{18}$eV in water whose minimum energy is $10^{9}$eV, respectively. It is clear from the figures that the width of the range distribution increases rapidly, as their primary energy increases. Also, as the primary energy decreases, the width of range distribution becomes narrower and approaches to a  delta function-type, the limit of which denotes no fluctuation. 
It is interesting that the range distribution can be well approximated as the normal distribution above $\sim 10^{14}$ eV where the total Cherenkov light yields comes almost from the accompanied electromagnetic cascade showers due to direct electron pair, bremsstrahlung, and photo nuclear interaction  initiated by the muon(see, Part2) and they are given as,
\begin{equation}
\label{PRE}                         
P\left(R;E_{0}\right)=\frac{1}{\sqrt{2\pi}\sigma}exp\left(-\frac{R-<R>}{2\sigma^{2}}\right),
\end{equation}
where $E_{0}$, $<R>$ and $\sigma$ are primary energy, the average value of ranges and the standard deviations, respectively.
Their average ranges, standard deviations and relative variances (standard deviations divided by averages) in water are given in Table \ref{tab:AVE}. Also, it is interesting that their relative variances decrease slightly as their primary energies increase.

In order to examine each characteristic of stochastic process, such as the bremsstrahlung, direct electron pair production and photo nuclear interaction, we compare the hypothetical range distribution in which a specified stochastic process only  is assumed to occur with the real distribution in which all stochastic processes occur as the result of the competition effect among these processes.
We compare the real range distributions with the hypothetical range distribution in Figures \ref{fig:RF13Sp} to \ref{fig:RF18Sp}.

\begin{table}[!t]                   
\begin{center}
\caption{The average values, the standard deviations and the relative variances of the range distributions of muons from $10^{11}$eV to $10^{18}$eV in water.}
\label{tab:AVE}
\scalebox{0.9}[0.9]{
\begin{tabular}{c|l|l|c}
\hline
\noalign{\smallskip}
$E_{0}$ [eV] & $<R>$ [km]        & $\sigma$ [km]        & $\sigma/<R>$         \\
\noalign{\smallskip}
\hline
$10^{11}$ & 3.56$\times 10^{-1}$ & 2.52$\times 10^{-2}$ & 7.07$\times 10^{-2}$ \\ \hline
$10^{12}$ & 2.43                 & 4.71$\times 10^{-1}$ & 1.94 $\times 10^{-1}$ \\ \hline
$10^{13}$ & 7.28                 & 2.02                 & 2.78$\times 10^{-1}$ \\ \hline
$10^{14}$ & 1.26$\times 10^{1} $ & 3.49                 & 2.77$\times 10^{-1}$ \\ \hline
$10^{15}$ & 1.78$\times 10^{1} $ & 4.57                 & 2.57$\times 10^{-1}$ \\ \hline
$10^{16}$ & 2.30$\times 10^{1} $ & 5.41                 & 2.36$\times 10^{-1}$ \\ \hline
$10^{17}$ & 2.79$\times 10^{1} $ & 6.14                 & 2.20$\times 10^{-1}$ \\ \hline
$10^{18}$ & 3.29$\times 10^{1} $ & 6.81                 & 2.07$\times 10^{-1}$ \\ \hline
\noalign{\smallskip}
\end{tabular}
}
\end{center}
\end{table}

 Here, the symbol \textit{d} in these figures means the hypothetical range distribution in which only direct electron pair production is taken into account and the bremsstrahlung and nuclear interaction are neglected. The symbols  \textit{b} and \textit{n} have similar meaning to that of \textit{d}. The symbol \textit{t} means the range distribution in which all interactions are taken into account(The ture distribution).
 From the shapes of the distributions and their maximum frequencies for different stochastic processes in Figures \ref{fig:RF13Sp} to \ref{fig:RF18Sp}, it is clear that energy losses in the direct electron  pair production are of small fluctuation, while both the bremsstrahlung and photo nuclear interaction are of bigger fluctuation and the fluctuation in photo nuclear interaction becomes remarkable when compared with bremsstrahlung as primary energy increases.
%
\subsubsection{The diversity of individual muon behavior}
\begin{figure*}[!t]
\begin{center}
\resizebox{1.0\textwidth}{!}{\includegraphics{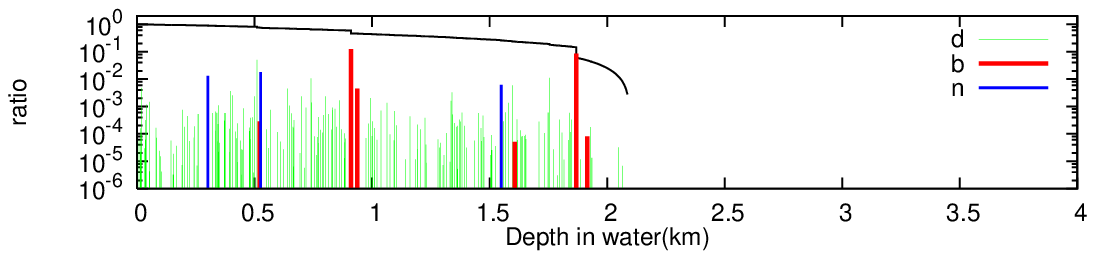}}	       
\caption{The energy losses with the shortest range for $10^{12}$eV muons.}
\label{fig:ELS12}
\resizebox{1.0\textwidth}{!}{\includegraphics{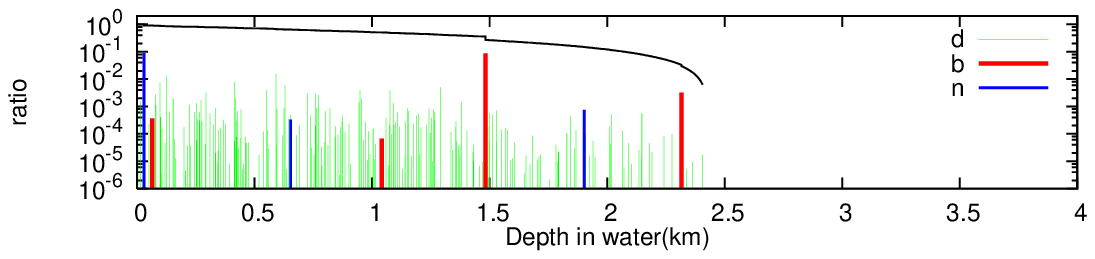}}	      
\caption{The energy losses with the average-like range for $10^{12}$eV muons.}
\label{fig:ELA12}
\resizebox{1.0\textwidth}{!}{\includegraphics{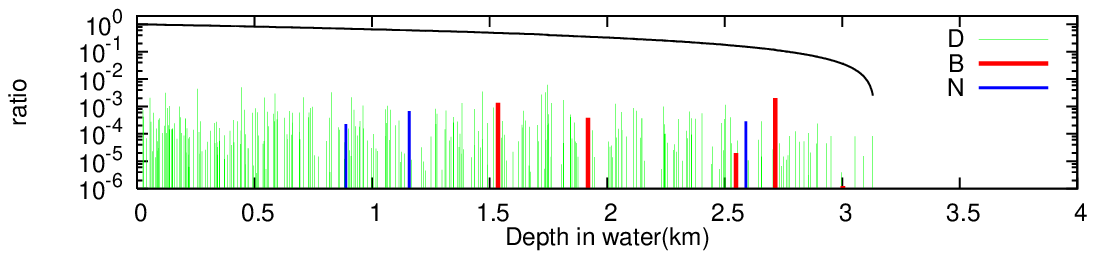}}	      
\caption{The energy losses with the longest range for $10^{12}$ eV muons.
(For interpretation of the references to color in legend of these Figures 17 to 20, the reader is referred to the web version of this article.)}
\label{fig:ELL12}
\end{center}
\end{figure*}
\begin{figure*}[!t]
\begin{center}
\resizebox{1.0\textwidth}{!}{\includegraphics{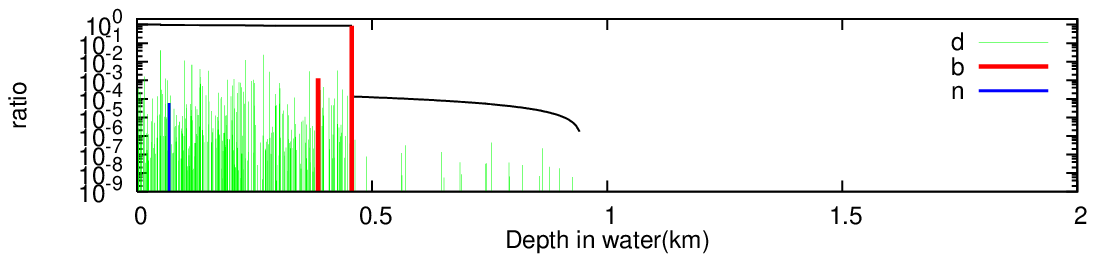}}	  
\caption{The energy losses with the shortest range for $10^{15}$eV muons. The figure is a magnification of Figure \ref{fig:ELS15}}
\label{fig:ELS15_2km}
\resizebox{1.0\textwidth}{!}{\includegraphics{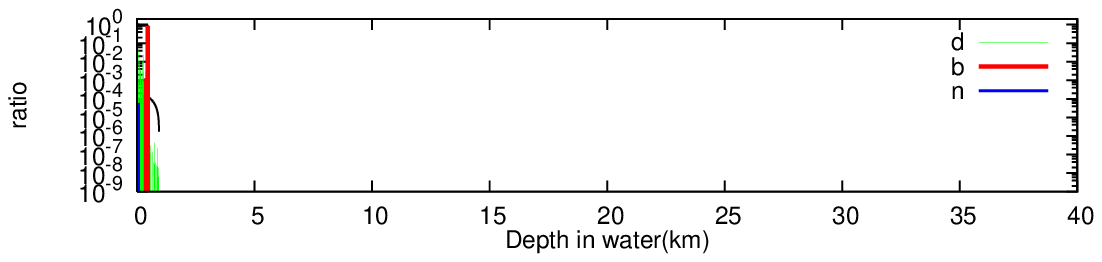}}	      
\caption{The energy losses with the shortest range for $10^{15}$eV muons.}
\label{fig:ELS15}
\resizebox{1.0\textwidth}{!}{\includegraphics{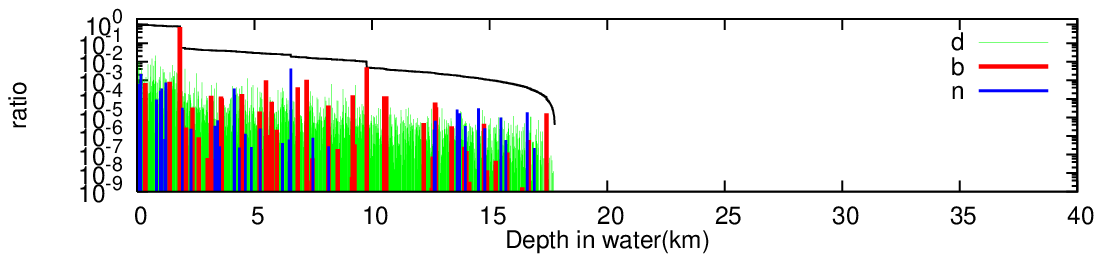}}	      
\caption{The energy losses with the average-like range for $10^{15}$eV muons.}
\label{fig:ELA15}
\resizebox{1.0\textwidth}{!}{\includegraphics{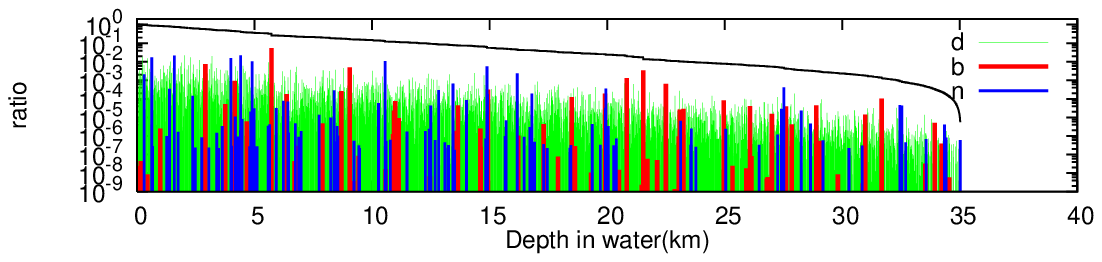}}	      
\caption{The energy losses with the longest range for $10^{15}$eV muons.
(For interpretation of the references to color in legend of these Figures 21 to 24, the reader is referred to the web version of this article.)}
\label{fig:ELL15}
\end{center}
\end{figure*}
\begin{figure*}[!t]
\begin{center}
\resizebox{1.0\textwidth}{!}{\includegraphics{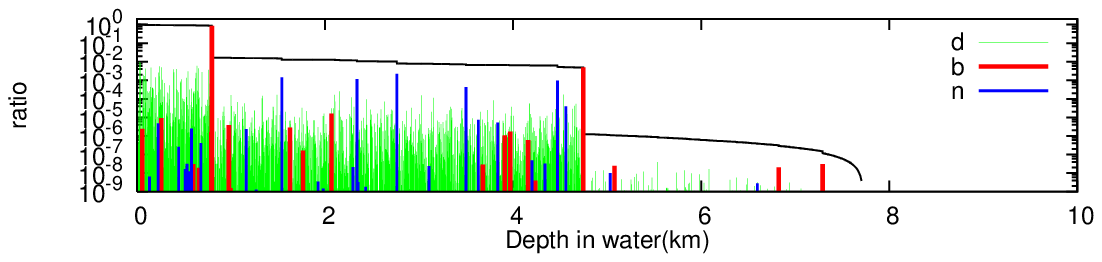}}	
\caption{The energy losses with the shortest range for $10^{18}$eV muons. The figure is a magnification of Figure \ref{fig:ELS18}.}
\label{fig:ELS18_10km}
\resizebox{1.0\textwidth}{!}{\includegraphics{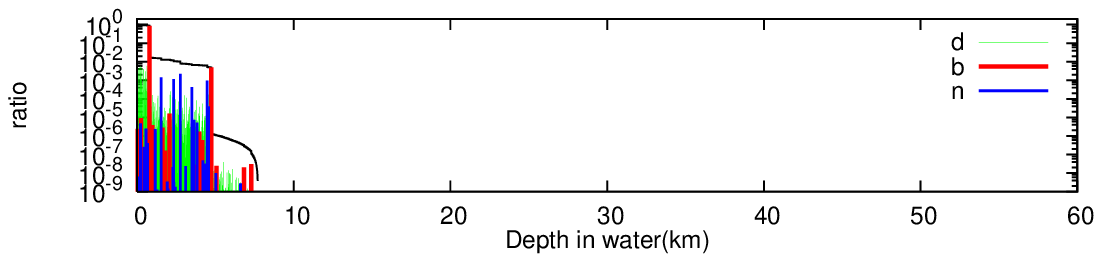}}	
\caption{The energy losses with the shortest range for $10^{18}$eV muons.}
\label{fig:ELS18}
\resizebox{1.0\textwidth}{!}{\includegraphics{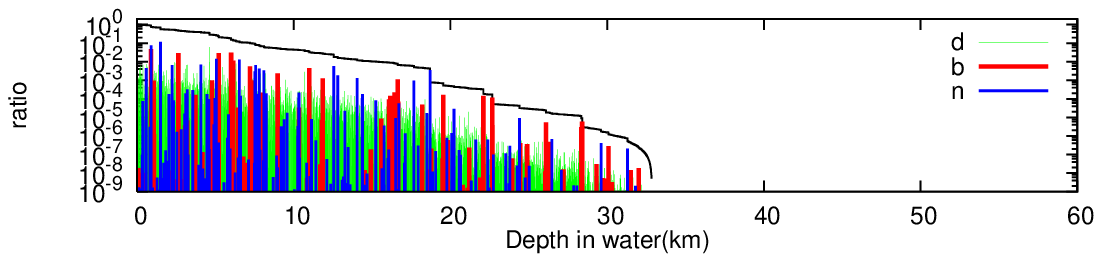}}	
\caption{The energy losses with the average-like range for $10^{18}$eV muons.}
\label{fig:ELA18}
\resizebox{1.0\textwidth}{!}{\includegraphics{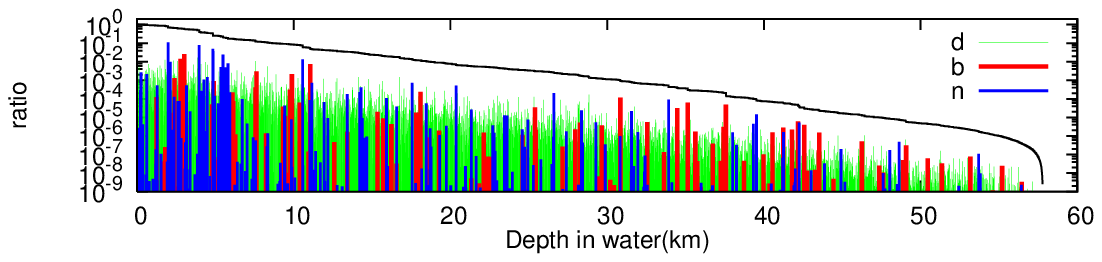}}	
\caption{The energy losses with the longest range for $10^{18}$ eV muons.
(For interpretation of the references to color in legend of these Figures 25 to 28, the reader is referred to the web version of this article.)}
\label{fig:ELL18}
\end{center}
\end{figure*}
In Figures \ref{fig:ELS12} to \ref{fig:ELL18}, we show the diversities of  muon behaviors for the same primary muon energies with regard to their ranges (or their energy losses). Also, in Table \ref{tab:ELT}, we summarize the characteristics of the events in Figure \ref{fig:ELS12} to \ref{fig:ELL18}. 
In these figures, all interaction points due to the processes of bremsstrahlung, direct electron pair production and photo nuclear interaction and all energy losses at respective points due to these processes are recorded.

In order to clarify the diversities among the real range distributions (or real energy loss distributions), we show the muons with the shortest range, the muons with longest range and the muons with average like range. The average-like range is defined as the nearest range to the average range which is calculated from the total sampled ranges.

In Figures \ref{fig:ELS12} to \ref{fig:ELL12}, we give the characteristic behaviors with the shortest range, the average-like range and the longest range as a function of the energy loss and the depths traversed for the same energy of $10^{12}$ eV in water\footnote{As for the characteristic behaviors of high energy muons which are shown Figures \ref{fig:ELS12} to \ref{fig:ELL12}, we suggest the readers that they look at the pictures with colors in the WEB page.}. 
In the Figures \ref{fig:ELS12} to \ref{fig:ELL12}, we utilize the same scale in depth to clarify the diverse behaviors by the same incident energies, namely, those with the shortest range, with the average like-range and with the longest range, respectively.
 In figures, the abscissa denotes the depths where the interactions concerned occur. The 'needles' with different colors denote the energy losses due to direct electron pair (green, d), bremsstrahlung (red, b) and photo nuclear interaction (blue, n), respectively. The ordinate shows the ratio of the energy loss concerned to the incident energy. Namely, 'needles' denote the energy losses due to the interaction concerned and line graphs (black curve) denote the ratios of the muon energies at the depths concerned to their incident energies. The abrupt changes in them denote the catastrophic energy losses for muons (see Figures \ref{fig:ELS15_2km} and \ref{fig:ELS18_10km}).

\begin{table*}[!t]              
\begin{center}
\caption{The details of the characteristics on the muons with the shortest range, the average-like range ,the longest range and the average range.}
\label{tab:ELT}
\scalebox{0.9}[0.85]{
\begin{tabular}{c|c|c|c|c|c|c|c}
\hline
\noalign{\smallskip}
	& Range & Energy loss & Number of   & Energy loss    & Number of   & Energy loss & Number of \\
$E_{0}=10^{12}eV$	&	[km]	& by brems  & interaction & by direct pair & interaction & by nuclear  & interaction \\
\noalign{\smallskip}
\hline
$<$Average$>$  &  2.43                &  1.10$\times 10^{11}$ & 4.74     & 1.57$\times 10^{11}$ & 243                 & 4.54$\times 10^{10}$ & 3.44\\ \hline
Shortest       &  4.65$\times 10^{-2}$&  2.15$\times 10^{11}$ & 6        & 1.52$\times 10^{11}$ & 208                 & 3.72$\times 10^{10}$ & 3 \\ \hline
Average-like   &  2.43                &  8.97$\times 10^{10}$ & 4        & 1.34$\times 10^{11}$ & 221                 & 8.86$\times 10^{10}$ & 3 \\ \hline
Longest        &  3.14                &  3.80$\times 10^{9}$  & 5        & 1.04$\times 10^{11}$ & 299                 & 1.19$\times 10^{9}$  & 3 \\ \hline
\noalign{\smallskip}
$E_{0}=10^{15}eV$\\ 
\noalign{\smallskip}
\hline
$<$Average$>$  & 1.78$\times 10^{1}$  & 3.53$\times 10^{14}$ & 48.1      & 4.74$\times 10^{14}$ & 6.80$\times 10^{3}$ & 1.67$\times 10^{14}$ &  55\\ \hline
Shortest       & 9.44$\times 10^{-1}$ & 8.66$\times 10^{14}$ &  2        & 1.34$\times 10^{14}$ &   367               & 5.90$\times 10^{10}$ &   1\\ \hline
Average-like   & 1.78$\times 10^{1}$  & 7.50$\times 10^{14}$ & 49        & 2.35$\times 10^{14}$ &  5489               & 9.31$\times 10^{12}$ &  37\\ \hline
Longest        & 3.50$\times 10^{1}$  & 7.53$\times 10^{13}$ & 71        & 8.02$\times 10^{14}$ & 13722               & 1.11$\times 10^{14}$ & 105\\ \hline
\noalign{\smallskip}
$E_{0}=10^{18}eV$\\ 
\noalign{\smallskip}
\hline
$<$Average$>$ & 3.28$\times 10^{1}$ & 3.37$\times 10^{17}$ & 108         & 4.39$\times 10^{17}$ & 2.57$\times 10^{4}$ & 2.25$\times 10^{17}$ & 172\\ \hline
Shortest      & 7.72$\times 10^{0}$ & 8.75$\times 10^{17}$ &  28         & 1.19$\times 10^{17}$ &  5760               & 6.23$\times 10^{15}$ &  40\\ \hline
Average-like  & 3.28$\times 10^{1}$ & 1.68$\times 10^{17}$ & 118         & 5.58$\times 10^{17}$ & 29321               & 2.74$\times 10^{17}$ & 196\\ \hline
Longest       & 5.78$\times 10^{1}$ & 5.71$\times 10^{16}$ & 162         & 5.77$\times 10^{17}$ & 46542               & 3.66$\times 10^{17}$ & 277\\ \hline
\noalign{\smallskip}
\end{tabular}
}
\end{center}
\end{table*}

It is seen from figures and Table \ref{tab:ELT} that there is not so big difference between the case with the shortest range and one with the longest range. In the case with the shortest range (Figure \ref{fig:ELS12}), we find two catastrophic energy losses (at $\sim$910 meters and $\sim$1870 meters) due to bremsstrahlung play the important role in the range. In the case with average-like range (Figure \ref{fig:ELA12}), we can find one catastrophic energy loss due to bremsstrahlung at $\sim$1.48 kilometer. However, in the case of the longest range (Figure \ref{fig:ELL12}), we cannot find the catastrophic energy losses due to bremsstrahlung and, instead, we can find that almost energy losses are due to many number($\sim$300) of direct electron pair production events with smaller energies. It is seen from Figure \ref{fig:RF131518} that the fluctuation effect in the range distribution is not so effective for $10^{12}$ eV muon and their different behaviors with the different ranges are not so impressive.

Figure \ref{fig:ELS15_2km} shows the same in Figure \ref{fig:ELS15} in different scale. The shortest range, $\sim$940 meter (Figure \ref{fig:ELS15_2km}), is far shorter compared with the longest one, $\sim$35.0 kilometers (Figure \ref{fig:ELL15}). It is seen from Figure \ref{fig:ELS15_2km} and the Tables that bremsstrahlung plays a decisive role as the cause of catastrophic energy loss, ($\sim$96.5\% of total energy at $\sim$450 meters). 86.6\% of the total energy is lost by 2 bremsstrahlungs, 13.4\% by 367 direct electron pair productions and $5.9 \times 10^{-3}$\% by 1 photo nuclear interaction. In Figure \ref{fig:ELL15}, we give the case for the longest range. Here, large numbers of direct electron pair production with rather small energy loss play an important role,  as shown similarly in Figure \ref{fig:ELL12}. Here, 80.2\% of the total energy is lost by 13722 direct electron pair productions, 7.53\% by 71 bremsstrahlungs and 11.1\% by 105 photo nuclear interactions. In Figure \ref{fig:ELA15}, we give the case with the average like range.  Here, 23.5\% of the total energy is lost by 5489 direct electron pair productions, 23.5\% by 49 bremsstrahlungs and 0.93\% by 37 photo nuclear interactions, while in the real averages, 47.4\% of the total energy is lost by 6800 direct electron pair productions, 35.3\% by 48.1 bremsstrahlung and 16.7\% by 55.0 photo nuclear interactions.
\begin{table}[h]            
\begin{center}
\caption{The Ratios of energies transferred from bremsstrahlung, direct electron pair production and photo nuclear interaction to the total energy loss.}
\label{tab:ratio}
\scalebox{0.85}[0.9]{
\begin{tabular}{c|c|c|c}
\hline 
$E_{0}=10^{12}eV$ & Brems & Direct Pair & Nuclear\\ 
\hline
$<$Average$>$&3.37$\times 10^{-1}$&5.26$\times 10^{-1}$&1.37$\times 10^{-1}$\\ \hline
Shortest     &5.32$\times 10^{-1}$&3.76$\times 10^{-1}$&9.20$\times 10^{-2}$\\ \hline
Average-like &2.87$\times 10^{-1}$&4.30$\times 10^{-1}$&2.83$\times 10^{-1}$\\ \hline
Longest      &3.50$\times 10^{-2}$&9.54$\times 10^{-1}$&1.10$\times 10^{-2}$\\ \hline 
$E_{0}=10^{15}eV$\\ 
\hline
$<$Average$>$&3.40$\times 10^{-1}$&4.98$\times 10^{-1}$&1.62$\times 10^{-1}$\\ \hline
Shortest     &8.66$\times 10^{-1}$&1.34$\times 10^{-1}$&5.90$\times 10^{-5}$\\ \hline
Average-like &7.54$\times 10^{-1}$&2.37$\times 10^{-1}$&9.36$\times 10^{-3}$\\ \hline
Longest      &7.62$\times 10^{-2}$&8.11$\times 10^{-1}$&1.12$\times 10^{-1}$\\ \hline
$E_{0}=10^{18}eV$\\ 
\hline
$<$Average$>$&3.24$\times 10^{-1}$&4.59$\times 10^{-1}$&2.17$\times 10^{-1}$\\ \hline
Shortest     &8.75$\times 10^{-1}$&1.19$\times 10^{-1}$&6.23$\times 10^{-3}$\\ \hline
Average-like &1.68$\times 10^{-1}$&5.58$\times 10^{-1}$&2.74$\times 10^{-1}$\\ \hline
Longest      &5.71$\times 10^{-2}$&5.77$\times 10^{-1}$&3.66$\times 10^{-1}$\\ \hline
\end{tabular}
}
\end{center}
\end{table}

 In Figures \ref{fig:ELS18_10km} to \ref{fig:ELL18},  we show the similar relations for $10^{18}$eV muons as shown in Figures \ref{fig:ELS15_2km} to \ref{fig:ELL15}. The case with shortest range in Figure \ref{fig:ELS18_10km} (Figure \ref{fig:ELS18}) has a strong contrast to that with the longest range. The manner of the energy loss in Figure \ref{fig:ELS18_10km} is drastic with two big catastrophic energy losses due to bremsstrahlung($\sim$0.8 km, $\sim$4.7 km), while that in Figure \ref{fig:ELL18} is moderate with no catastrophic energy loss. The shortest range , $\sim$7.7 kilometers, is far shorter compared with the longest range, $\sim$57.8 kilometers. It is seen from Figure \ref{fig:ELS18_10km} and the Tables that bremsstrahlung plays a decisive role as the cause of catastrophic energy loss. 87.5\% of the total energy is lost by 28 bremsstrahlungs, 11.9\% by 5760 direct electron pair productions and 0.623\% by 40 photo nuclear interactions. In Figure \ref{fig:ELL18}, we give the case with the longest range. Here, 57.7\% of the total energy is lost by 46542 direct electron pair productions, 36.6\% by 277 photo nuclear interactions and only 5.71\% by 162 bremsstrahlungs in the complete absence of catastrophic energy losses. In Figure \ref{fig:ELA18}, we give the case average-like range. Here, 55.8\% of the total energy is lost by direct electron pair production, 16.8 \% by 118 bremsstrahlung and 27.4\% by 196 photo nuclear interactions, while, in the real averages, 43.9\% of the total energy is lost by $2.57\times10^{4}$ direct electron pair productions, 33.7\% by 108 bremsstrahlung and 22.5\% by 172 photo nuclear interactions. Thus, it is can be concluded that the diversity among muon propagation with the same primary energy should be noticed.

In Table \ref{tab:ratio}, we give the proportion in the energy losses among bremsstrahlung, direct electron pair production and photo nuclear interaction for different ranges. It is easily understood from the table that in the cases with the shortest range, we obtain the proportions occupied by the bremsstrahlung $5.32 \times 10^{-1}$, $8.66 \times  10^{-1}$ and $8.75 \times 10^{-1}$ for $10^{12}$ eV, $10^{15}$ eV, and $10^{18}$ eV, respectively, while in the cases with the longest ranges, we obtain the proportions occupied by direct electron pair production, $9.54 \times 10^{-1}$, $8.11 \times 10^{-1}$, $5.77 \times 10^{-1}$, respectively. Namely, catastrophic energy losses by bremsstrahlung make decrease their ranges remarkably, while the direct electron pair production has high probability to lose smaller energy and this effect makes it possible their ranges to be extended. Also, it is seen from the information around <average> that, on the average, a half of the total energy loss is occupied by the direct electron the energy loss and the ratio of the energy loss due to bremsstrahlung which is the origin of the strong fluctuation should be considered carefully.
%
\section{Conclusion}
\label{sec:4}
Many authors have examined the survival probability on high energy muons by the Monte Carlo simulation which we call \textit{the $v_{cut}$ procedure}, expecting to utilize their results for the analysis of the muon neutrino events in the KM3 detector in future. We have developed \textit{the time sequential procedure} in the Monte Carlo simulation without introducing $v_{cut}$, by which we give the same results with those by \textit{the $v_{cut}$ procedure} as far as the muon energy behavior is concerned.

Using our method, we have examined the diversity on the muons' ranges (or, energy losses) in detail.
Also, we have clarified that range energy distribution is well presented by the normal distribution over $10^{14}$ eV to $10^{18}$ eV and we have examined the degree on the fluctuation among three stochastic processes, by using the hypothetical range distributions due to the stochastic characters.

Since we are only interested in behaviors of the high energy muon, there is no essential differences between the results obtained by \textit{the $v_{cut}$ procedure} and \textit{the time sequential procedure}. 
However, the difference in the results obtained from both \textit{the time sequential procedure} and \textit{the $v_{cut}$ procedure} may appear in Cherenkov light yields. As the Cherenkov light yields are closely related to the energy estimation of the muon concerned, the difference should be carefully examined. In a subsequent paper (Part2)\cite{Okumura}, we will discuss Cherenkov light yields due to high energy muon in greater detail with regard to muon straggling.

\section*{Acknowledgments}
Authors are grateful for the help of Bryan Pattison for correcting English of the manuscript.

%
\appendix
\setcounter{figure}{0}
\section{The Monte Carlo procedure for \textit{the time sequential procedure}}
\label{App}
\begin{figure}[!t]       
\begin{center}
\resizebox{0.45\textwidth}{!}{\includegraphics{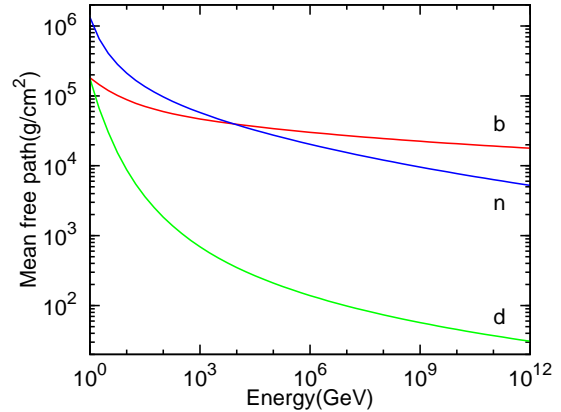}}
\caption{Mean free paths for bremsstrahlung (Curve b), direct electron pair production (Curve d), photo nuclear interaction (Curve n).}
\label{fig:MFP}
\end{center}
\end{figure}
If the differential cross section for bremsstrahlung\cite{Kelner}, direct pair production\cite{Kokoulin} and photo nuclear interaction\cite{Borg} are denoted by $\sigma_{b}\left(E,E_{\gamma}\right)dE_{\gamma}$, $\sigma_{d}\left(E,E_{e}\right)dE_{e}$ and $\sigma_{n}\left(E,E_{n}\right)dE_{n}$, respectively.  
Then, the mean free paths of the muons due to bremsstrahlung, direct electron pair production and photo nuclear interaction are given respectively as
\begin{equation}        
\lambda_{b}\left(E\right)= \frac{1}{\frac{N}{A}\int_{E_{\gamma_{min}}}^{E_{\gamma_{max}}} \sigma_{b}\left(E,E_{\gamma}\right)dE_{\gamma}}\\
\end{equation}
\begin{equation}        
\lambda_{d}\left(E\right)= \frac{1}{\frac{N}{A}\int_{E_{e_{min}}}^{E_{e_{max}}} \sigma_{d}\left(E,E_{e}\right)dE_{e}}
\end{equation}
\begin{equation}        
\lambda_{n}\left(E\right)= \frac{1}{\frac{N}{A}\int_{E_{n_{min}}}^{E_{n_{max}}} \sigma_{n}\left(E,E_{n}\right)dE_{n}}
\end{equation}
 Also, the resultant mean free path for these radiative processes are given as
\begin{equation}        
\frac{1}{\lambda_{total}\left(E\right)}= \frac{1}{\lambda_{b}\left(E\right)}+\frac{1}{\lambda_{d}\left(E\right)}+\frac{1}{\lambda_{n}\left(E\right)}
\end{equation}
The integrations for (A.1) to (A.3) are performed over kinematically allowable ranges. In (A.1), $E_{\gamma, min}$ is taken to be satisfied with such a condition that $E_{\gamma, min}/E_{0}$ is sufficiently smaller than $E_{min}/E_{0}$, where $E_{0}$ and $E_{min}$ denote the primary energy of the muon and the minimum energy of the muon for observation.  $E_{min}$ is taken as 1 GeV throughout present paper. In Figure A.1, we give the mean free paths for bremsstrahlung, direct electron pair production and photo nuclear interaction are given as the function of the primary energy. 
\begin{figure}[!t]
\begin{center}
\resizebox{0.48\textwidth}{!}{\includegraphics{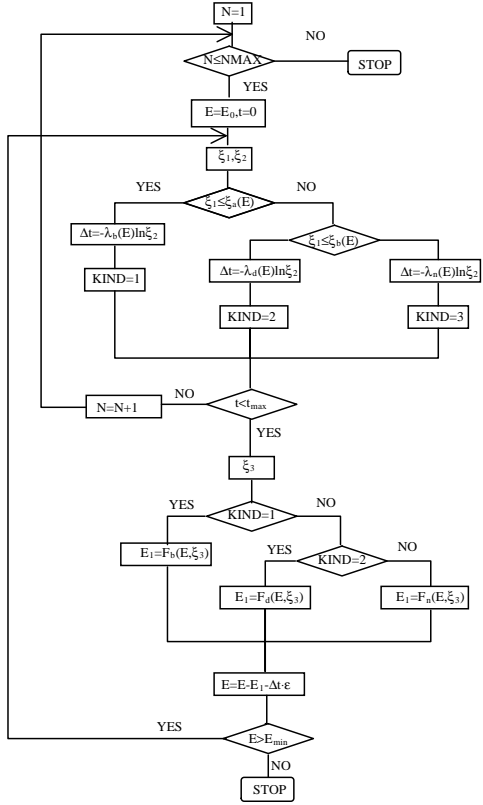}} 
\caption{Flow chart of muon propagation.}
\label{Flow} 
\end{center}
\end{figure}

The most important procedures in our Monte Carlo method are only two. The first one is where the interaction occurs. For  bremsstrahlung, the traversed distance for the interaction is determined with the use of $\xi$, a uniform random number between (0,1), as follows.
\begin{equation}        
\Delta t = -\lambda_{b}\left(E\right)log\xi,
\end{equation}
Similarly, for direct electron pair production,
\begin{equation}        
\Delta t = -\lambda_{d}\left(E\right)log\xi,
\end{equation}
Similarly, for the photo nuclear interaction,
\begin{equation}        
\Delta t = -\lambda_{n}\left(E\right)log\xi,
\end{equation}
In our Monte Carlo simulation, the energy losses due to each stochastic  process are sampled by the following equations with the use of $\xi$, the uniform random number, between $\left(0,1\right).$   For bremsstrahlung
\begin{equation}        
\label{xi_b}
\xi = \frac{\int_{E_{\gamma_{min}}}^{E_{\gamma}} \sigma_{b}\left(E,E_{\gamma}\right)dE_{\gamma}}{\int_{E_{\gamma_{min}}}^{E_{\gamma_{max}}} \sigma_{b}\left(E,E_{\gamma}\right)dE_{\gamma}},
\end{equation}
 From (\ref{xi_b}),
\begin{equation}        
E_{\gamma} = F_{b}\left(E,\xi\right),
\end{equation}
 For direct electron pair production
\begin{equation}        
\label{xi_d}
\xi = \frac{\int_{E_{e_{min}}}^{E_{e}} \sigma_{d}\left(E,E_{e}\right)dE_{e}}{\int_{E_{e_{min}}}^{E_{e_{max}}} \sigma_{d}\left(E,E_{e}\right)dE_{e}},
\end{equation}
 From (\ref{xi_d}),
\begin{equation}        
E_{e} = F_{d}\left(E,\xi\right),
\end{equation}
 For photo nuclear interaction
\begin{equation}        
\label{xi_n}
\xi = \frac{\int_{E_{n_{min}}}^{E_{n}} \sigma_{n}\left(E,E_{n}\right)dE_{n}}{\int_{E_{n_{min}}}^{E_{n_{max}}} \sigma_{n}\left(E,E_{n}\right)dE_{n}},
\end{equation}
 From (\ref{xi_n}),
\begin{equation}        
E_{n} = F_{n}\left(E,\xi\right),
\end{equation}
Exactly speaking, the samplings in equations (A.8), (A.10) and (A.12) are carried out in $v$, the ratios of the sampled energies to their primary energies, not the sampled energies themselves.
 For the discrimination among stochastic processes in our Monte Carlo simulation let us introduce the following equations.
\[
\xi_{a}\left(E\right) = \frac{1/\lambda_{b}\left(E\right)}{1/\lambda_{b}\left(E\right)+1/\lambda_{d}\left(E\right)+1/\lambda_{n}\left(E\right)}
\]
\[
\xi_{b}\left(E\right) = \frac{1/\lambda_{b}\left(E\right)+1/\lambda_{d}\left(E\right)}{1/\lambda_{b}\left(E\right)+1/\lambda_{d}\left(E\right)+1/\lambda_{n}\left(E\right)}
\]
In Figure \ref{Flow}, we give a flow chart for our Monte Carlo simulation.
\begin{figure}[ht]           
\begin{center}
\resizebox{0.45\textwidth}{!}{\includegraphics{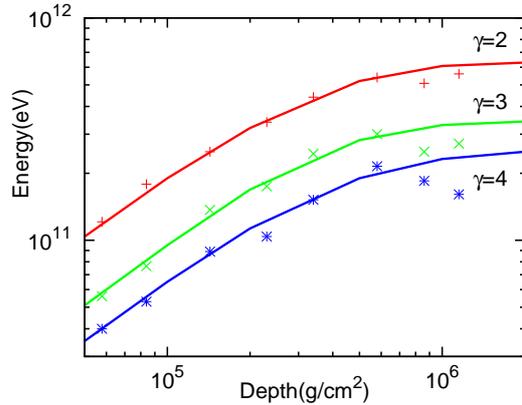}}
\caption{The average energies of the muons. The lines denote our, while symbols denote Misaki and Nishimura.}
\label{fig}
\end{center}
\end{figure}
The validity of our Monte Carlo procedure by \textit{the time sequential procedure} should be carefully examined in the two methods which are independent of each other. Namely, the first one is the comparison of our procedure with the analytical theory which is explained in the present Appendix and the other is the comparison of our procedure (\textit{the time sequential procedure}) with the different kind of Monte Carlo procedure (\textit{the $v_{cut}$ procedure}) which is mentioned in the text. Particularly, it is  best that the results obtained by a Monte Carlo procedure are checked by the procedures which are  methodologically independent of the Monte Carlo procedure, reaching the same results. In Figure A.3, the average energies of the muons are given as a function of the depths under the preposition of muon energy spectrum at sea level with indexes 2, 3 and 4, obtained by \textit{the time sequential procedure} \cite{Takahashi} \cite{Takahashi2} and they are compared with results obtained by the analytical theory based on the Nishimura-Kamata formalism in the cascade shower theory \cite{Kiraly} and the agreement between them is very good, which surely guarantees the validity of our \textit{time sequential procedure}.
%
\bibliography{<your-bib-database>}

\begin{thebibliography}{00}

\bibitem{IceCube}       
F.Halzen and S.R.Klein, Rev.Sci.Instrum. \textbf{81} (2010) 081101

\bibitem{Antarctic}     
ANTARES Collb., arXiv:1007.1777v1 [astro-ph.HE] 11 Juli 2010

\bibitem{Mediterranean} 
NEMO Collb.,Astropart.Phys.\textbf{33}, (2010) 263

\bibitem{Baikal}        
Baikal.Collb.,Astropart.Phys.\textbf{7}, (1997) 263

\bibitem{Mando}         
Mando and Sona, Nuovo Cimento,\textbf{10},(1953) 1275

\bibitem{Zatsepin}      
G.T.Zatsepin and E.D.Michaelchi, Jour.Phys. Soc.Japan \textbf{17}, Suppl.A lll, (1962) 356 

\bibitem{Hayman}        
P.J.Hayman, N.S.Palmer and A.W.Wolfendale,  \textbf{275A}, (1963) 391

\bibitem{Nishimura}     
J.Nishimura: Proc.Int.Conf.on Cosmic Rays.Vol.6 (Jaipur,1964) 224

\bibitem{Miyake}        
S.Miyake,V.S.Narashimham and P.V.Ramana Murthy, Nuovo Cimento,\textbf{32}, (1964) 1524

\bibitem{Kobayakawa}    
K.Kobayakawa, Nuovo Cimento,\textbf{B47}, (1967) 156

\bibitem{Kiraly}        
E.Kiraly, P.Kiraly and J.L.Osborn, J.Phys.A,5 (1972) 444

\bibitem{Misaki}        
A.Misaki and J.Nishimura, Uchusen Kenkyuu \textbf{21},(1976) 250 \\
   (ICR-Rerort-4577-4, University of Tokyo(1977))

\bibitem{Gurentsov}     
V.I.Gurentsov, G.T.Zatsepon and E.D.Mikhalchi, Sov.J.Nucl.Phys.\textbf{23}, (1976) 527

\bibitem{Minorikawa}    
Y.Minorikawa,T.Kitamura and K.Kobayakawa, Nuovo Cimento, \textbf{C4}, (1981) 471

\bibitem{Bhattacharyya} 
D.P.Bhattacharyya, Nuovo Cimento, \textbf{C9}, (1986) 404

\bibitem{Allkofer}      
O.C.Allkofer and D.P.Bhattacharyya, Phys.Rev.\textbf{D34}, (1986) 1368

\bibitem{Lipari}        
P.Lipari, T.Stanev, Physical Review D \textbf{44}, (1991) 3543

\bibitem{Antonioli}     
P.Antonioli et al.,Astroparticle Physics \textbf{7}, (1997) 357

\bibitem{Dutta}         
S.Iyer Dutta, M.H.Reno, I.Sarcevic and D.Seckel Phys.Rev D \textbf{63}, (2001) 094020

\bibitem{Klimushin}     
S.I.Klimushin, E.V.Bugaev, I.A.Sokalski, Physical Review D \textbf{64}, (2001) 014016\\
I.A.Sokalski, E.V.Bugaev, S.I.Klimushin, Physical Review D \textbf{64}, (2001) 074015

\bibitem{Chirkin}       
D.Chirkin, W.Rhode, hep-ph \textbf{0407075v2} (2008)

\bibitem{Kudryavtsev}   
V.A.Kudryavtsev, Computer Physics Communications \textbf{180}, (2009) 339

\bibitem{Bottai}        
S.Bottai and L.Perrone, Nuclear Instruments and Methods in Research A \textbf{459}, (2001) 319-325

\bibitem{Takahashi}     
N.Takahashi, A.Misaki, A.Adachi, N.Ogita, Y.Okamoto, K.Mitsui, H.Kujirai, S.Miono, O.Saavedra, Proc.18th ICRC Bangalore India \textbf{11}, (1983) 443 

\bibitem{Nishimura2}       
J.Nishimura, Handbuch der Physik, XLVI/2, 'COSMIC RAYS II', Springer-Verlag, (1967) 1

\bibitem{Takahashi2}       
N.Takahashi et al, Uchusen-Kenkyu, \textbf{28}, (1984) 120

\bibitem{Okumura}           
Y.Okumura, N.Takahashi and A.Misaki, A subsequent paper after the present paper.

\bibitem{Kelner}            
S.R.Kelner, R.P.Kokoulin, A.A.Petrukhin, Preprint MEPhI 024-95 Moscow, (1995) ; CERN SCAN-9510048

\bibitem{Kokoulin}          
R.P.Kokoulin, A.A.Petrukhin, Proc.11st ICRC Budapest \textbf{MU-41}, (1969) 277

\bibitem{Borg}              
V.V.Borg, A.A.Petrukhin, Proc.14th ICRC Munchen \textbf{6}, (1975) 1949
\end{thebibliography}

\end{document}